\documentclass[sigconf]{acmart}

\renewcommand\footnotetextcopyrightpermission[1]{} % removes footnote
\AtBeginDocument{%
  }

\begin{document}

\title{Resonant and Stochastic Vibration in Neurorehabilitation}


\author{Ava Hays}
\affiliation{%
  \institution{The University of Michigan}
  \city{Ann Arbor, MI}
  \country{USA}}
\email{avahays@umich.edu}

\author{Nolan Kosnic}
\affiliation{%
  \institution{The University of Michigan}
  \city{Ann Arbor, MI}
  \country{USA}}
\email{nkosnic@umich.edu}

\author{Ryan Miller}
\affiliation{%
  \institution{The University of Michigan}
  \city{Ann Arbor, MI}
  \country{USA}}
\email{ryanmilr@umich.edu}

\author{Kunal Siddhawar}
\affiliation{%
  \institution{The University of Michigan}
  \city{Ann Arbor, MI}
  \country{USA}}
\email{siddhawr@umich.edu}


\begin{abstract}
Neurological injuries and age-related decline often impair the processing of sensory information and disrupt motor coordination, gait, and balance. As neuroplasticity has become better understood, vibration-based interventions have emerged as a promising means to stimulate both sensory pathways and motor circuits to promote functional recovery. This paper reviews stochastic and resonant vibration, examines their mechanisms, therapeutic rationales, and clinical applications. We synthesize evidence on whole-body vibration for balance, mobility, and fine motor recovery in aging adults, stroke survivors, and individuals with Parkinson’s disease, highlighting challenges in parameter optimization, generalized efficacy, and safety. We also evaluate recent advances in focused muscle vibration and wearable stochastic resonance devices for upper-limb rehabilitation, identifying their clinical potential as well as limitations in scale, ecological validity, and standardization. Across these modalities, we outline the key variables shaping therapeutic outcomes and summarize ongoing efforts to refine protocols, enhance usability, and integrate vibration therapies into broader neurorehabilitation frameworks. This survey concludes by identifying the most pressing research needs for translating vibration-based interventions into reliable, deployable clinical tools.
\end{abstract}

\received{08 December 2025}
%%\received[revised]{date}
%%\received[accepted]{date}

\maketitle

\section{Introduction}
\label{sec:section}
Vibration-based interventions like whole body vibration and focused muscle vibration offer methods to improve neuromuscular function, sensory processing, and recovery. This review outlines the underlying principles of these interventions, stochastic and resonant vibrations, then details how they are implemented for the human body. Then the clinical applications of these methods are examined in their effectiveness and challenges. Finally, potential solutions for individual applications are discussed as well as emerging work in the field. 

\section{Background}
\label{sec:section}
Traumatic brain injuries, strokes, and to a lesser extent aging, can reduce a person's ability to integrate new sensory information, which can disrupt coordination and balance \cite{albert2012neurorehabilitation}. Up until the latter half of the 20th century, it was widely believed that brain capabilities lost due to brain injury could not be regained once full brain development was reached \cite{albert2012neurorehabilitation}. However, since then neuroplasticity has become a widely studied process, which is defined as the brain reorganizing its structure and functionality in response to stimuli to change activity levels. This process can be split up into either functional reorganization, in which existing neurons change roles, or synaptic plasticity, in which neurons form new connections \cite{puderbaugh2023neuroplasticity}. 
\par
Since it has been shown that the improved functionality that comes from neurorehabilitation improve participation and quality of life of those affected \cite{khan2017neurorehabilitation}, understanding the variables which increase ability for neuroplasticity have become paramount. Through this motivation, it has been understood that 1) behavioral and motor experience (action and repetition), and 2) incorporating sensory feedback, enhance plasticity \cite{nudo2013recovery, maier2019principles}. Therefore, finding ways to induce motor function to an individual, and to feed these sensory signals to the brain, is the motivation for the methods discussed throughout this paper.

\subsection{Stochastic Vibrations}
\label{sec:subsection}
% White et al = white2019promise
% McDonnell et al = mcdonnell2011benefits
% Matthews et al = matthews2024stochastic
% Moss et al = moss2004stochastic
Stochastic vibration (or stochastic resonance) is a mechanism where adding random variability like noise can help nonlinear systems, like the human nervous system, enhance the detection or processing of weak signals \cite{white2019promise, mcdonnell2011benefits, matthews2024stochastic, moss2004stochastic}. Generally noise is assumed to obstruct clarity and performance \cite{matthews2024stochastic, mcdonnell2011benefits}. Essentially the random noise combines with the weak signal enabling the resultant signal to pass the systems threshold for detection \cite{matthews2024stochastic}. However, this noise must be an optimal amount to enhance performance without degrading the signal \cite{moss2004stochastic, matthews2024stochastic, white2019promise}. 
\par
This works in neurorehabilitation in two ways. The constant randomized oscillations cause the neuromuscular system to make continuous posture adjustment, strengthening the system in a low-intensity and low-impact way. Additionally, the vibrations can enhance the brain's ability to detect sensory information, in order to adjust, by amplifying weak signals to above the activation threshold. This is particularly helpful to those with poor sensory feedback due to age or neurological conditions. 

\subsection{Resonant Vibrations}
\label{sec:subsection}
Resonant vibration refers to resonance which is matching a system's natural frequency to an external frequency which results in an enhanced output signal \cite{vincent2021vibrational}. Vibrational resonance occurs when a predicted  high-frequency harmonic force is introduced to enhance a system's response to a low-frequency signal \cite{vincent2021vibrational}. Every part of the body has a natural frequency it oscillates at \cite{vincent2021vibrational}. In resonant vibration, the external vibration matches these internal oscillations to amplify their movement. This increase in vibration for muscles and tendons activates muscle spindles which can improve muscle stiffness, joint stability, and proprioception which is a body’s sense of position \cite{needle2024neural, white2019promise}. Overall resonant vibration amplifies reflexive and neural responses in a predictable way.  
\par
Both stochastic and resonant vibrations aim to enhance neuromuscular and sensory function, but they do so through different mechanisms. Stochastic vibration uses random oscillations while resonant vibration uses tuned external frequencies to match natural frequencies in the body. 

\subsection{History \& Early Integration}
\label{sec:subsection}
The earliest use of vibrations for therapeutic benefit came in the mid 1960s. In what is now called Tonic Vibration Reflex (TVR), it was observed that muscles involuntarily contracted when its tendon was vibrated at certain high sinusoidal frequencies at low amplitude  \cite{stillman1970vibratory}. It was through this application of resonant vibrations that gave the initial justification for using this for neuromuscular applications rather than for comfort. 
\par
Stochastic resonance had its origins from physics and nonlinear systems theory, but found an application in neurorehabilitation when it was theorized that the “detections” SR hoped to perceive could be sensory input from muscles. The first application of this principle came from posture control and balance from greater sensory input from the feet and ankles, and produced considerable success which will be discussed in future sections.
\par
This foundational work identified the principles that have been built upon in the last few decades that are currently utilized in various methods and machines. The effectiveness of each of these applications, as well the challenges and remaining questions to maximize the success of these applications, will now be discussed.

%start of WBV
\section{Whole Body Vibration}
Whole body vibration (WBV) is a training or treatment that uses oscillating platforms to deliver mechanical vibrations to subjects while they sit, stand, or exercise \cite{van2021reporting}. The vibrations are characterized by frequency, magnitude, wave form, exposure time, and number of daily bouts \cite{van2021reporting, oroszi2020vibration}, these vibrations are either vertical or horizontal \cite{van2021reporting}. A specific type of whole body vibration which is seen in most clinical applications is stochastic resonance (SR-WBV) which randomizes the vibrations characteristics \cite{donocik2017alterations, elfering2013stochastic, herren2018effects}. 
\par
Overall, WBV combines muscle and skin components \cite{oroszi2020vibration}. The muscle component induces muscle contractions which essentially "exercises" the muscles leading to an increase in strength \cite{van2021reporting, chanou2012whole}. The skin component uses sensitive skin mechanoreceptors which detect the vibrations and send them to the brain through the spinal cord \cite{oroszi2020vibration} helping with sensory deficits \cite{donocik2017alterations}. In relation to the brain, studies have concluded through fMRI scans that SRT affects the basal ganglia loops in the brain that assist in gait and posture, \cite{kaut2016postural, kaut2016stochastic}. Therefore, WBV has been used in neurorehabilitation to aid in balance \cite{calderone2025exploring, chanou2012whole, costantino2018effects, donocik2017alterations, elfering2013stochastic, park2018comparison} and mitigate symptoms of neurological conditions \cite{calderone2025exploring, costantino2018effects, park2018comparison}. 

\subsection{Balance \& Mobility}
\label{sec:subsection}
One clinical application of whole body vibration is for improving balance and mobility for those with, often age related, musculoskeletal or metabolic conditions \cite{chanou2012whole}. As people age their muscle strength, postural control, and sensory perception decline making them higher risk for falls \cite{white2019promise}. WBV provides a low-impact, more practical form of exercise that is uniquely suited to elderly or frail subjects who are unable to engage in traditional strength training. Overall by stimulating muscle activity and increasing sensory feedback, WBV is emerging as a promising method of improving stability, lower-limb, function and confidence in older individuals.
\par
SR-WBV specifically uses small random vibrations sent throughout the body to improve postural stability and sensory control. The vibrations activate muscle spindles and other sensory receptors which helps the body's sense of position and movement which allows the nervous system to detect and correct to maintain balance \cite{donocik2017alterations}.
\par
A study was conducted using SR-WBV with 187 women from ages 19-74 over the course of six weeks and used posturography to track the change in a body’s center of pressure (COP) while standing. Essentially the more COP a subject has the larger they sway, and the harder it is for them to remain stable and balanced. This study found that after SR-WBV participants had a lower mean velocity of their COP and that their weight was more evenly distributed between legs \cite{donocik2017alterations}. Thus they stood firmer and corrected sway better.
\begin{figure}[h]
  \centering
  \includegraphics[width=\linewidth]{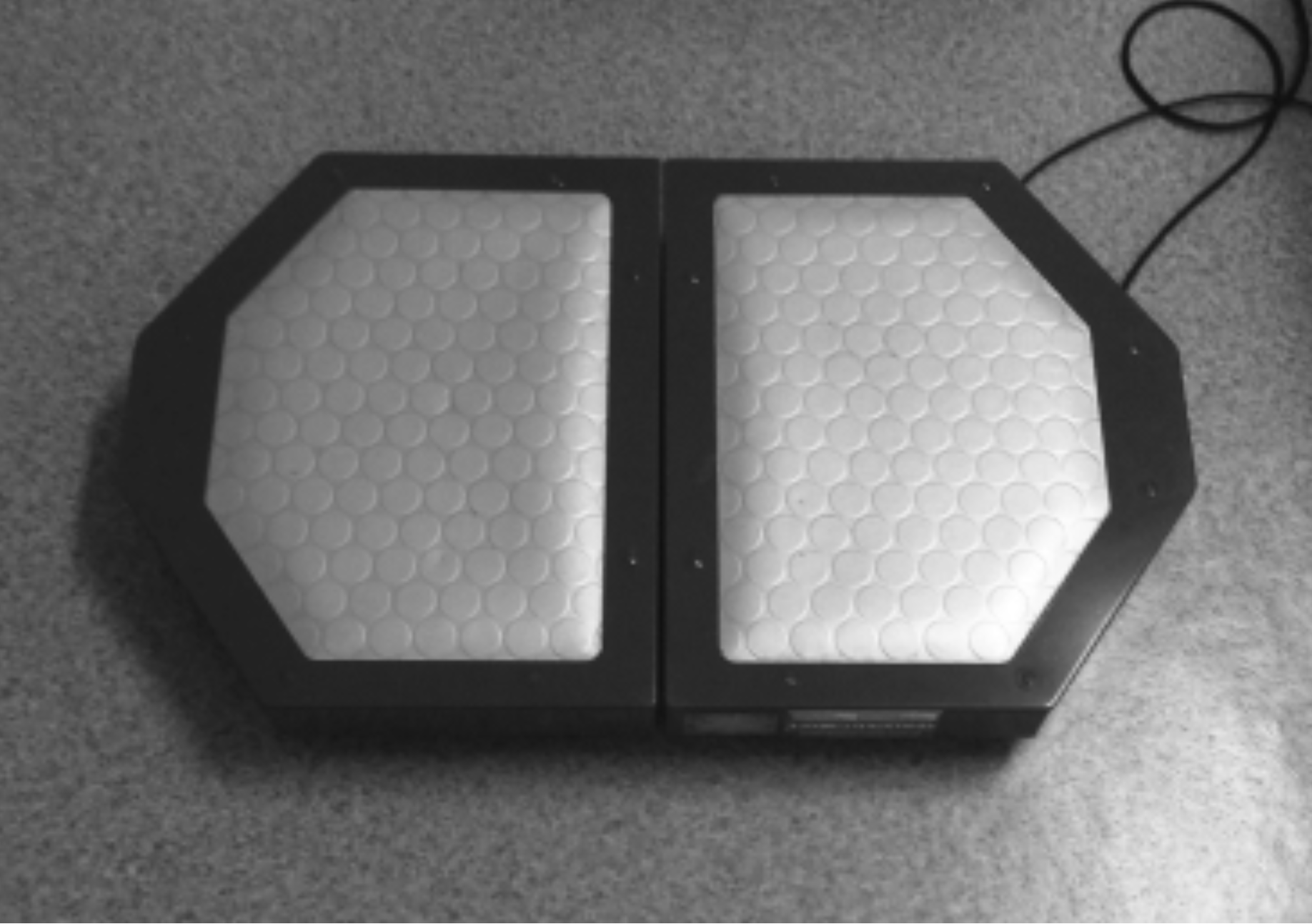}
  \caption{Double plate posturography device used to track COP \cite{donocik2017alterations}.}
  \Description{Unnecessary?}
\end{figure}
\par
Another study used SR-WBV for office workers to increase balance, confidence, and musculoskeletal health. Participants in this study performed better on balance tests after SR-WBV and reported feeling more stable on their feet. Additionally the study showed an improvement in musculoskeletal health and a decrease in musculoskeletal pain in those with lower back pain prior to the study. While this test was not specifically done on the elderly, it supported the usage of SR-WBV in balance and musculoskeletal health, which can be applied to the elderly as very few side effects were reported and there were no participant drop outs \cite{elfering2013stochastic}. 
\par
WBV can also enhance the effects of exercises that combine physical activity with mental challenges. For example in a study of care-home residents aged 79-98, SR-WBV (shown in Figure 2) was used in combination with exergame-dance training (shown in Figure 3) which requires participants to move their limbs in accordance with visual or auditory instructions. This works to increase muscle strength as well as reflexes in the elderly. This study found improvements in lower limb strength, coordination in legs and arms, better mental flexibility and executive function, suggesting that the small vibrations may improve communication between brain and muscles \cite{de2020combining}. 
\begin{figure}[h]
  \centering
  \includegraphics[width=\linewidth]{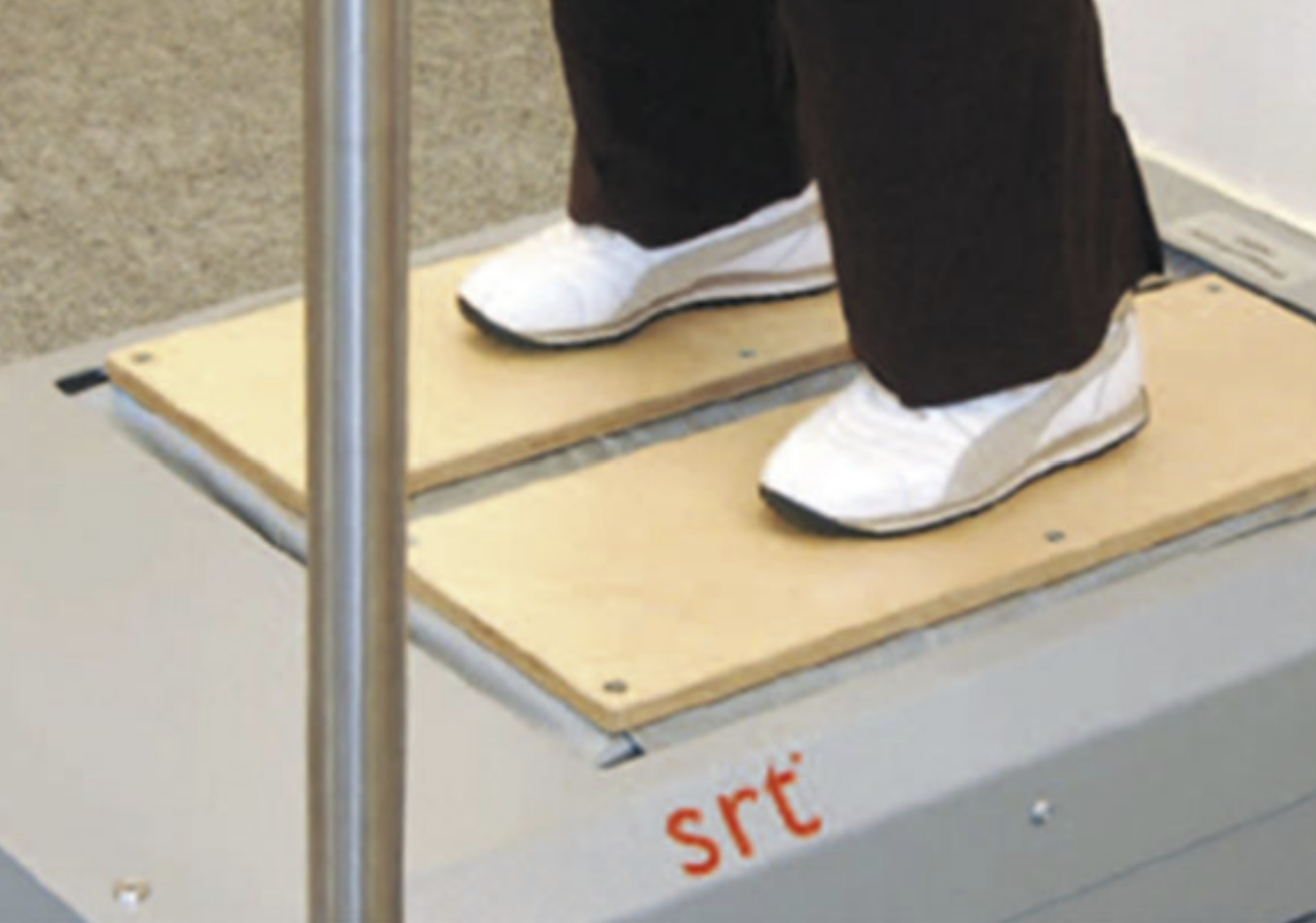}
  \caption{Platform used for vibration training sessions \cite{de2020combining}.}
  \Description{Unnecessary?}
\end{figure}
\par
\subsubsection{Balance \& Mobility Challenges \& Contributions}
\label{sec:subsubsection}
WBV is not yet used as a primary treatment to improve balance and mobility in elderly populations because its effectiveness depends heavily on optimizing several key parameters referred to as the Big Five: vibration amplitude, vibration frequency, method of application, session duration/frequency, and total intervention duration \cite{oroszi2020vibration}. 
\par
The scalability and broader clinical deployment of WBV are limited by its variable efficacy across factors such as age, BMI, neuromuscular status, and overall frailty, as well as by the increased health risks that vibration therapy may pose to vulnerable older adults. In the study by Donocik et al., SR-WBV was found to be more effective in younger, taller, and slimmer women, with a clear negative correlation between the index of balance improvement and both age and BMI. Although SR-WBV has demonstrated measurable improvements in balance among participants aged 60 years and older \cite{donocik2017alterations}, the wide age range sampled across different studies makes it difficult to determine the consistency of WBV’s effects in older adults specifically. Additionally, parameter optimization must account for the medical fragility of the target population, as conditions common in older adults may contraindicate or limit safe participation in WBV interventions \cite{donocik2017alterations}. Further, there is a large variability in protocol across studies making the overall effectiveness of WBV for balance and mobility hard to determine.
\par
Some work has been done to optimize protocol, which can potentially lead to more standardization. For example, studies utilizing WBV-plus-exercise showed that 68\% of measures had significant improvements, whereas WBV-only showed significant improvement in only 41\% of measures \cite{orr2015effect}. Research on the effects of session frequency has been conducted with a study showing training three times a week had better results in the Timed Up and Go (TUG) test compared to those training two times a week \cite{orr2015effect}. Other ways in which research has attempted to address parameter optimization is by specifically detailing their protocols for better comparison.
\begin{figure}[h]
  \centering
  \includegraphics[width=\linewidth]{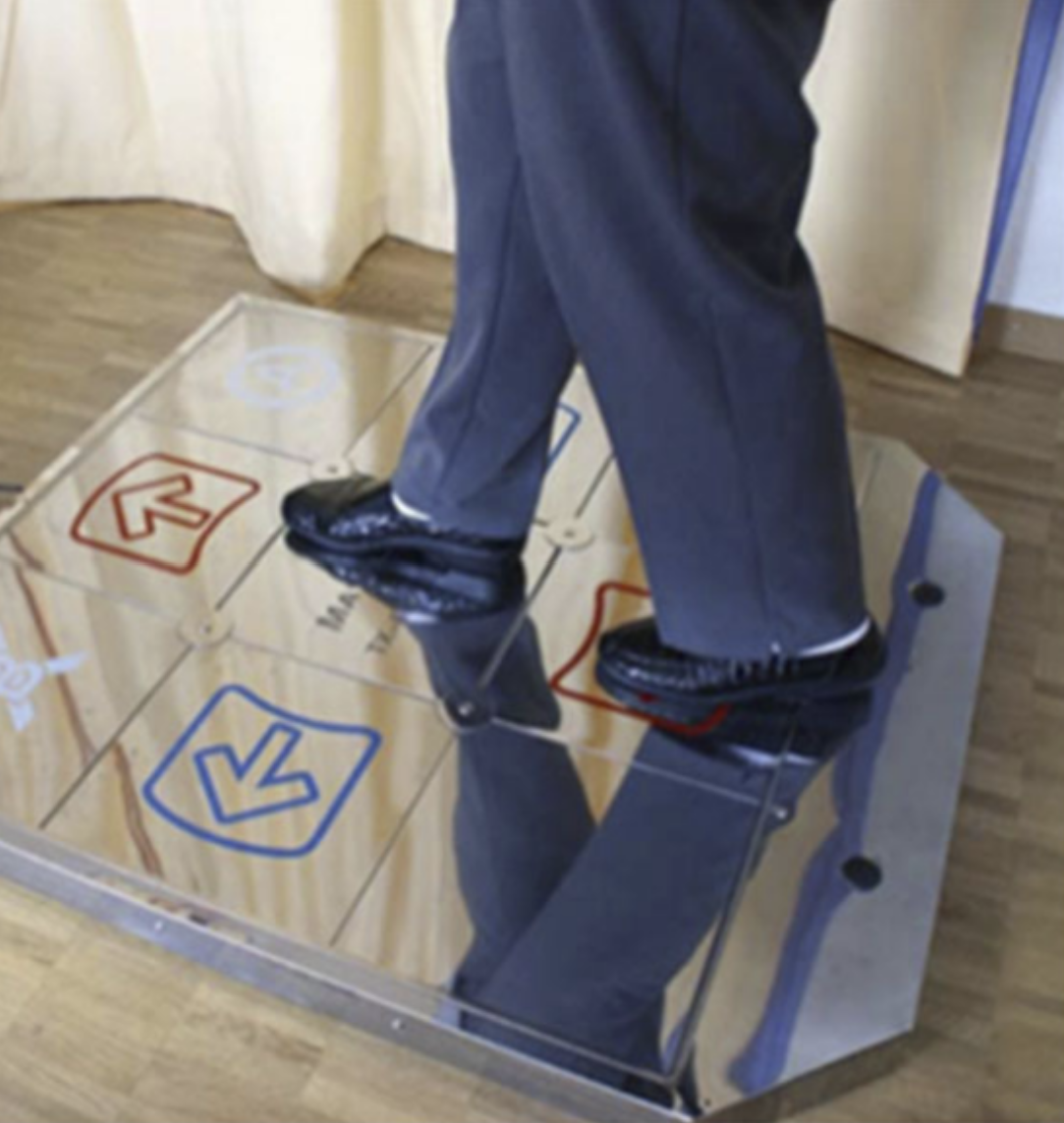}
  \caption{Virtual dance platform for training after vibration sessions \cite{de2020combining}.}
  \Description{Unnecessary?}
\end{figure}
\par
Further to address the changing efficacy for different groups, studies have shown that the most impaired groups, the frail and clinical patients, have a greater improvement when using WBV than healthy elderly individuals \cite{orr2015effect}. Additionally, to address the risk concerns for the elderly and frail a six-week static WBV study was conducted on nursing home residents found that subjects attended 96\% of the exercise sessions over the six weeks \cite{bautmans2005feasibility}, thus proving that WBV’s low-impact and low-volume nature can be tolerable for frail groups. Another way studies ensure safety is by strict exclusion criteria with many studies excluding cardiac disease, vestibular or neuromuscular disorders, neurological impairments \cite{orr2015effect}, insulin-dependent diabetes mellitus, heart pacemaker, or musculoskeletal disorders \cite{bautmans2005feasibility}.

\subsection{Fine Motor Skills}
\label{sec:subsection}
Another promising application of whole body vibration comes by targeting the neurological side effects of conditions which severely impair fine motor capabilities, such as side effects induced by strokes or Parkinson’s disease. These motor skills are more particular to extraordinary neurological conditions rather than simple aging, and therefore are measured separately in large clinical trials.
\par
 For example, in a large meta-analysis done by Park et al. \cite{park2018comparison}, various individual symptoms of stroke victims were studied. It was shown that of all neurological symptoms, spasticity was reduced the most. It offered immediate therapeutic relief in particular with ankle plantarflexion spasticity in chronic stroke patients \cite{chan2012effects}. This strongly relates to the previous section, as this spasticity is in direct relation to muscles corresponding to gait and balance. This paper, along with others\cite{fritton1997}, also cited an improvement in bone metabolism outcomes, which can also improve motor skills and mobility. However, there is skepticism as to the consistency of these results of improved spasticity. For example, Yang et al. found no significant improvement in mobility, listing inconclusive results.
\par
Additionally, the reduction of various motor skill disorders were studied in Parkinson’s patients. In a study with 56 PD participants, the experimental group receiving stochastic resonance therapy noted a 41.6\% improvement in rigidity, a 23.7\% improvement in bradykinesia, and a 30.8\% improvement in tremors \cite{kaut2016postural}. Although these improvements did not reach the level of significance in the between-group analysis due to the strong placebo response in PD studies (the control group saw a 14.5\% improvement in bradykinesia), they are significantly more consistent than non-stochastic (resonant) therapy \cite{arias2009effect}.

\subsubsection{Fine Motor Skills Challenges \& Contributions.}
\label{sec:subsubsection}
As we can see through a view of the literature, WBV has its greatest benefit to stroke patients through improvement of spasticity. However, it should be noted that, as far as symptom improvement goes, we are examining the spasticity that causes the balance issues separately from the balance issues themselves. According to Park et al. \cite{park2018comparison}, WBV was better at improving spasticity that impacts gait function rather than improving gait itself, which produced inconclusive results in another meta analysis \cite{costantino2018effects}. This is because spasticity is the thing which contributes to the overall dysfunction, however there are several more variables which contribute to gait. This makes it easier to identify parameters which improve ankle plantarflexion spasticity, for instance, but this does not generalize to tangible results (less falls) or improved functionality.
\par
In spite of this distinction, there is still more that needs to be done to further prove WBVs effectiveness in improving motor skills for the neurologically limited (including both stroke victims and Parkinson's patients). Park et al. \cite{park2018comparison} mentions that the meta analysis of stroke victims was too small, which implies increasing the study count of stroke victims. This research suffers from similar challenges to that which looks to improve mobility in the elderly such as the quality of these studies. For instance, more studies could be double blind procedures, include a control group, have larger sample sizes, or involve a long term follow up. 
\par
Different parameters could also be studied more extensively. As was mentioned previously, there is no standardization of the “Big Five” parameters. In the initial research, the greatest success in improving spasticity from WBV came from 12 Hz vibrations at 4 mm with a 2.3g acceleration force \cite{park2018comparison}, and the greatest success at improving motor skills for PD came from 4 treatments over 8 days at 7 Hz, 3 mm amplitude. Frequency selection and proper posture have also been identified to maximize muscular benefits \cite{yang2021effects}; however, all of these parameters are effective only for very particular muscular movements and cannot be generalized.

%Start of FMV
\section{Focused Muscle Vibration}
Whole‑body vibration (WBV), first explored as early as Charcot’s time, applies low‑frequency oscillations to the entire body but has shown limited effects because it must remain within a narrow frequency range to avoid adverse reactions. These limitations have been overcome by focused (local) vibration that delivers higher‑frequency stimulation directly to specific muscles or tendons, producing more targeted neuromuscular and clinical benefits in neurorehabilitation. Focal muscle vibration has historical roots dating back to Charcot and was further developed in the late 20th century as summarized in \cite{vigano2023focal}. 
\par
FMV uses a small mechanical transducer to deliver high-\ frequency oscillations directly to a specific muscle belly or tendon, usually in the range of about 90–300 Hz or more, which is much higher than typical whole‑body vibration frequencies. The device is placed perpendicular to the skin over the target musculotendinous area, producing tiny cyclic changes in muscle‑tendon length that strongly activate muscle spindle receptors and the tonic vibration reflex, leading to increased contractile activity and modulation of spinal and supraspinal motor circuits. These signals travel through fast myelinated afferent fibers and engage central motor centers, so FMV ends up acting like rapid, repeated small concentric–eccentric contractions, with measurable effects on strength, muscle activation patterns, and spasticity when applied in structured rehabilitation protocols (for example, several 10–30 minute sessions at set frequencies and rest intervals) \cite{saggini2017vibration, murillo2014focal}.

\subsection{Upper-limb Motion}
\label{sec:subsection} 
\par
One vibrotactile approach related to focal muscle vibration is stochastic resonance stimulation (SRS), in which low - amplitude mechanical noise is delivered to the skin to improve the detection of weak sensory signals in non - linear neural systems. Lynn et al. \cite{lynn2023effects} evaluated this concept in a randomized, sham-controlled crossover pilot study of sixteen children aged 3–16 years with hemiplegic cerebral palsy (MACS I–III). These children wore lightweight wrist and upper -arm bands containing piezoelectric actuators that delivered vibrotactile SRS or sham (devices off) while completing the Box and Block Test and Shriners Hospital Upper Extremity Evaluation (SHUEE) in a single laboratory session. Subthreshold SRS was individually titrated to about 80–90\% of each child’s detection threshold, and in a subsequent open-label phase a subset received above -threshold SRS at roughly 110–120\% of threshold, with blinded raters scoring SHUEE videos. Figures in the original article depict the soft wraps housing circular SRS discs on the wrist and upper arm during tabletop tasks, as well as the mobile application interface used to pair devices and adjust stimulation intensity, highlighting the fully
wearable, child-compatible setup.
\par
In the subthreshold randomized comparison, children moved on average approximately 1.8 more blocks per minute on the Box and Block Test with SRS than with sham (p$\approx$0.08, small effect size), and their SHUEE Spontaneous Functional Analysis scores increased by about 3 points (p<0.002), while Dynamic Positional Analysis scores rose by roughly 2.7 points without reaching statistical significance. In the open-label subset above the threshold, children transferred approximately 3.9 additional blocks per minute (p<0.001, moderate effect size) and showed mean gains of roughly 4.5 points in SHUEE-SFA (p$\approx$0.08) and 10.5 points in SHUEE-DPA (p$\approx$0.01), with no adverse events reported, suggesting that SRS is feasible and potentially beneficial to improve unimanual and bimanual hand function in this pediatric population.
\par
A similar technique has been tested in adults post-stroke using the TheraBracelet \cite{seo2019therabracelet}, a wrist-worn device delivering subthreshold random-frequency vibration to the paretic wrist during intensive task-practice therapy. In a triple-blinded pilot randomized controlled trial, twelve chronic stroke survivors were assigned to active or sham groups and completed 2-hour upper-extremity training sessions three times per week for two weeks while wearing identical wrist units, with the treatment group receiving imperceptible stimulation at 60\% of sensory threshold and the sham group receiving no stimulation. Figure 4 shows the compact tactor attached at the volar wrist beneath the sleeve during functional manipulation tasks, indicating no obstruction to movement. The intervention was feasible and well tolerated, with no adverse events or desensitization. On the Box and Block Test, the active group improved by about six additional blocks compared with baseline while the sham group showed no meaningful change, yielding a significant treatment-by-time interaction and a large effect size; gains partially persisted at 19-day follow-up, and Wolf Motor Function Test scores demonstrated favorable but non-significant trends with moderate-to-large effect sizes. Together with pediatric findings from vibrotactile stochastic resonance stimulation (where children with hemiplegic cerebral palsy demonstrated modest gains in Box and Block performance and SHUEE scores under subthreshold SRS, and larger improvements in an above-threshold open-label phase) these results highlight how lightweight wrist- and arm-based vibrotactile stimulation systems can augment upper-limb performance across neurologic populations while minimally interfering with real-world task execution.
\begin{figure}[h]
  \centering
  \includegraphics[width=\linewidth]{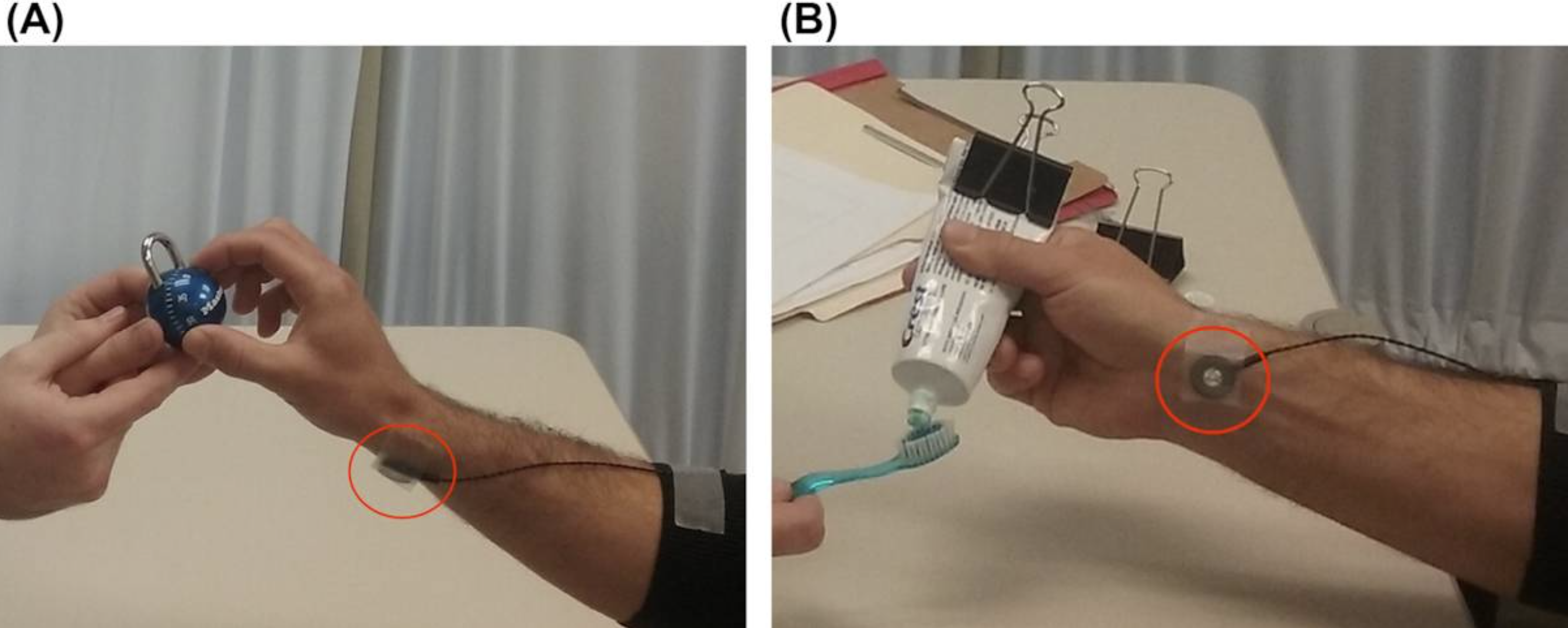}
  \caption{Placement of the TheraBracelet during task-practice therapy addressing hand object manipulation, such as opening a lock (A) and applying toothpaste (B). \cite{seo2019therabracelet}.}
  \Description{Unnecessary?}
\end{figure}

\subsubsection{Upper-limb Motion Challenges \& Contributions}
\label{sec:subsubsection}
Across both pediatric SRS and adult TheraBracelet trials, a central challenge is the limited scale and ecological reach of current evidence: samples are small, interventions are brief, and testing often occurs in highly controlled environments with constrained task sets (for example, single‑session protocols in children and 2‑week programs focused on tabletop dexterity in adults). These constraints mirror broader difficulties in FMV‑related wearable development, where devices remain wired, non‑waterproof, or clinic‑bound, complicating use in daily life and limiting long‑term adherence and outcome assessment. Practical issues such as imperfect blinding, missing data in younger children, and heterogeneous stroke/CP presentations further reduce statistical power and generalizability, making it hard to derive robust dose–response relationships, optimal stimulation parameters, or clear indications for specific impairment profiles. Collectively, these limitations help explain why, despite promising prototypes (TheraBracelet \cite{seo2019therabracelet}, VTS glove \cite{seim2021wearable}, other vibrotactile garments), there are still no widely adopted commercial FMV or vibrotactile neurorehabilitation products for upper‑limb function.
\par
At the same time, these studies make important contributions by showing that subthreshold or low‑level vibrotactile stimulation can be integrated with functional tasks in a wearable format and can yield clinically meaningful improvements in hand use. The TheraBracelet trial demonstrates that imperceptible, random‑\ frequency wrist stimulation can be delivered concurrently with intensive task‑practice in a triple‑blinded RCT, producing large effect sizes on Box and Block and favorable trends on Wolf Motor Function outcomes compared with equally dosed therapy alone. The pediatric SRS work shows that a single session of wrist/arm noise stimulation can acutely enhance spontaneous use and joint positioning of the impaired hand, with above‑threshold stimulation sometimes outperforming subthreshold noise, thereby challenging canonical assumptions about stochastic resonance dosing. Together with related vibrotactile wearables such as the VTS glove \cite{seim2021wearable}, these trials collectively position wearable FMV‑related stimulation as a feasible, low‑burden sensory adjunct that can be layered onto conventional therapy or daily activities, and they define clear next steps (larger, longer, multi‑site trials, better parameter optimization, and more user‑friendly hardware) for translating these concepts into scalable clinical technologies.

\subsection{Spasticity}
\label{sec:subsection} 
Focused muscle vibration has also been a growing trend for treating muscle spasticity, with several journals identifying its possible benefits for improving muscle strength and efficiency \cite{fattorini2023effectiveness}. Main approaches to treat spasticity utilize resonant FMV, as specific sinusoidal frequencies directly activate la afferents, the sensory fibers that detect muscle stretch. Activation of afferent projections to alpha motor neurons results in a phenomenon called the tonic vibration reflex (TVR), which causes muscle contraction and antagonist muscle relaxation \cite{murillo2014focal, saggini2017vibration}. While this approach is the most common for focused muscle vibration, which targets spinal reflex pathways directly, stochastic vibration is also used to activate weak sensory signals to surpass neural thresholds. This technique is primarily used to improve sensory and proprioceptive acuity and motor output, indirectly rather than relying on reflex modulation \cite{wiesenfeld1995stochastic, mendez2012improved}.

\subsubsection{Resonant Applications}
\label{sec:subsubsection}
Resonant studies typically use specific frequencies over varying durations, often lasting several days or weeks, with considerable diversity in parameters selected \cite{casale2014localized, cariati2021dose, caliandro2012focal, marconi2011long, tavernese2013segmental}. Most frequency modulation vibration (FMV) protocols are centered around 100 Hz, a potential resonant frequency for muscle spindle activation \cite{casale2014localized, caliandro2012focal, medica2020localized, toscano2019short}. However, lower frequencies, like the 30 Hz protocol, also show benefits \cite{cariati2021dose}. Other research indicates that vibrations can influence mechanistic pathways in the body. For instance, Caliandro found that FMV alters stretch-reflex behavior, demonstrating some spinal-level modulation, and other studies suggest that long-term cortical adaptations occur following repeated sessions \cite{caliandro2012focal, marconi2011long}. Functional improvements have also been documented during task-based paradigms, such as changes in reaching performance observed when vibration was applied during movement \cite{tavernese2013segmental}.
\begin{figure}[h]
  \centering
  \includegraphics[width=\linewidth]{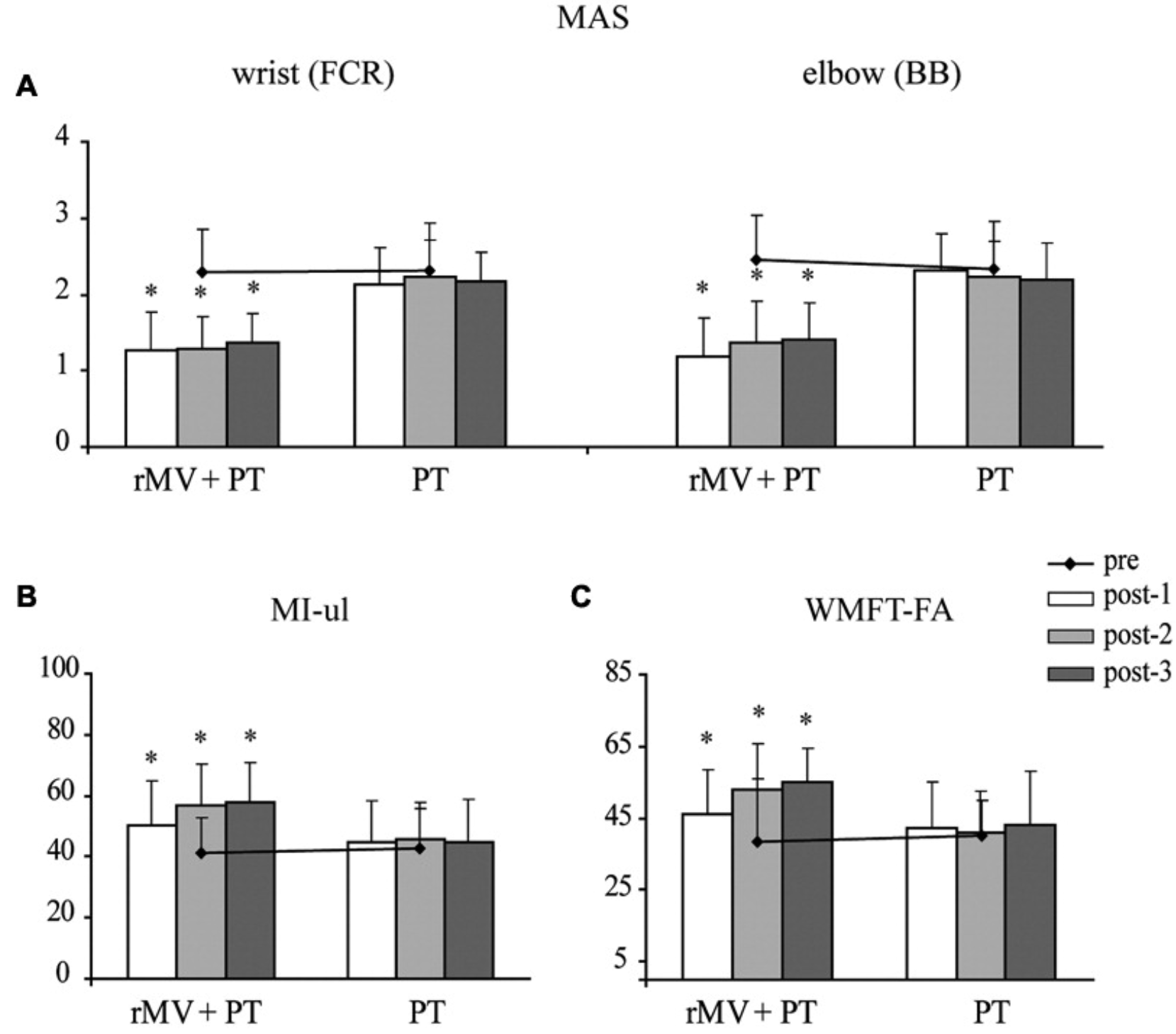}
  \caption{Results from a FMV experiment which show long-term cortical adaptations occur following repeated sessions\cite{marconi2011long}.}
  \Description{Unnecessary?}
\end{figure}

\subsubsection{Resonant Challenges \& Contributions}
\label{sec:subsection:subsubsection}
Studies involving resonant FMV techniques vary widely in their implementation. Across systematic reviews of the field that investigate the effects on over 850 people, there is broad agreement that although these studies share common goals, they vary widely in amplitude (0.2–2 mm), session duration, stimulation site, and treatment frequency, leaving no standardized protocol \cite{medica2020localized, zeng2023effects}. Commonly, studies used 100Hz for their vibration therapy, indicating that frequency is consistently being proven to improve muscle spasticity in the short term \cite{medica2020localized, caliandro2012focal, casale2014localized, toscano2019short}. However, it is important to recognize that, on its own, each of these studies has a small sample size. Still, FMV consistently produces short-term reductions on the Modified Ashworth Scale (MAS), though long-term retention remains scarce \cite{zeng2023effects}. 

\subsubsection{Stochastic Application} 
\par 
Stochastic-vibration studies almost always rely on wearable, subsensory stimulation, where broadband noise is applied below the perceptual threshold to enhance weak sensory inputs \cite{fallon2004stochastic, mildren2017frequency, sacco2018effects}. Because this approach aims to boost sensory fidelity rather than directly provoke reflexes, most implementations take the form of compact wrist or hand devices, small surface stimulators, or therapy-integrated bands.
\par
The wearables are mainly united by the technique they use to treat (subsensory noise), but vary notably in how they target different aspects of sensorimotor function. Some devices deliver wrist-based subsensory stimulation to enhance hand function remotely, showing that SR can improve distal performance without directly stimulating the involved muscle group \cite{Seo2014RemoteNoise}. Others apply vibrotactile noise directly to the hand to produce rapid gains in dexterity, emphasizing immediate performance effects rather than long-term adaptation \cite{nobusako2019subthreshold}. Therapy-integrated bands introduce continuous subsensory noise during rehabilitation, aiming to prime sensorimotor learning across repeated training sessions rather than focusing on single-task improvements \cite{vatinno2022using}. 
\begin{figure}[h]
  \centering
  \includegraphics[width=\linewidth]{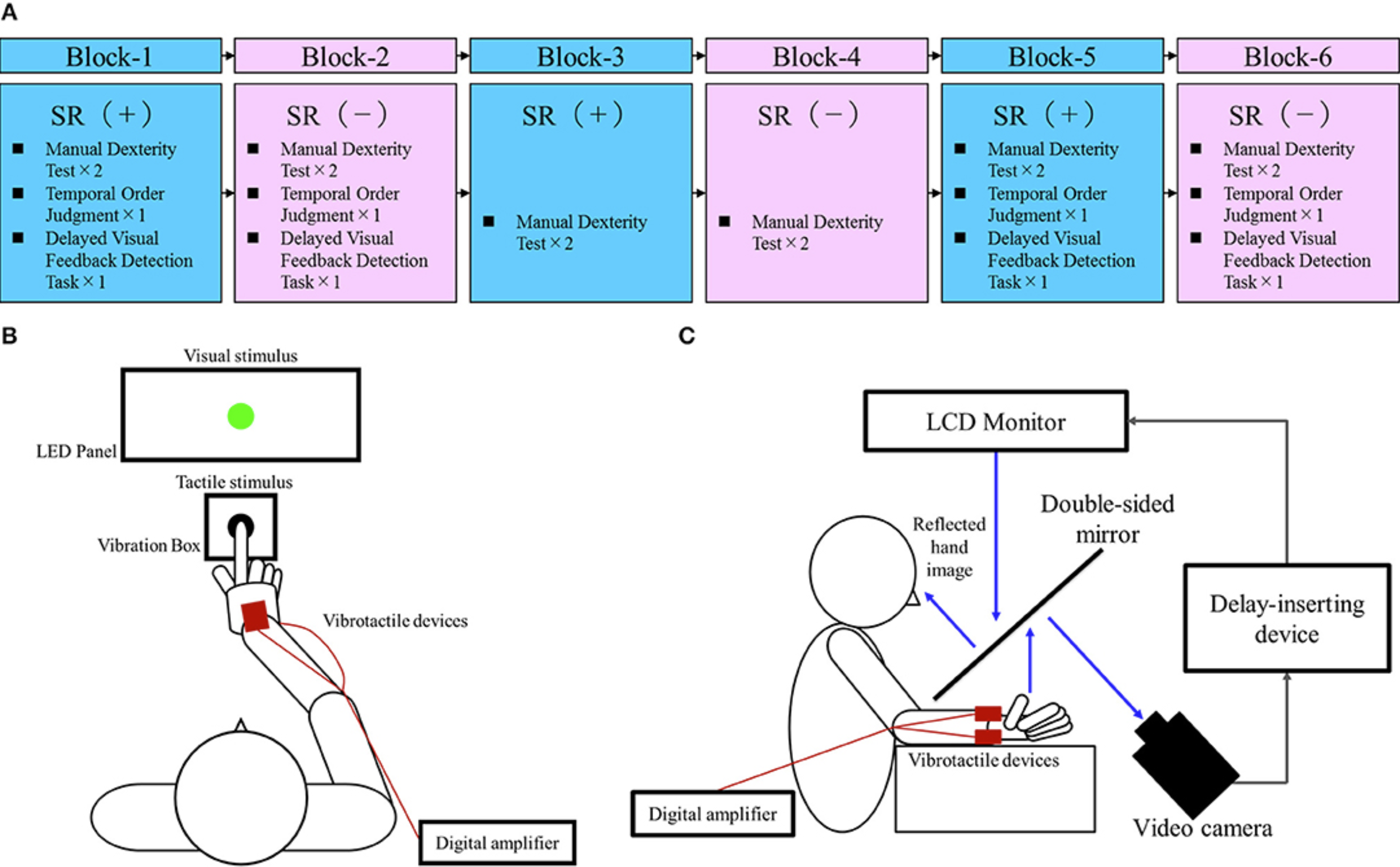}
  \caption{An outline of an experiment where children completed a temporal order judgment task using controlled visuo-tactile stimuli and a delayed visual feedback task, in which they viewed a time-shifted video reflection of their own hand. \cite{nobusako2019subthreshold}.}
  \Description{Unnecessary?}
\end{figure}
\par
\subsubsection{Stochastic Challenges \& Contributions}
\label{sec:subsection:subsubsection} 
\par
Across the wearable-focused literature, stochastic studies consistently emphasize sensation, dexterity, and motor-learning enhancement rather than direct spasticity reduction, and they place less specific frequency requirements than resonant LMV \cite{zeng2023effects, stein2010stochastic, seim2023relief}. This suggests that SR-based vibration may serve as a complementary strategy to therapy, providing sensory and motor benefits that could indirectly influence hypertonia. Stochastic applications are once again limited by the small sample sizes, but this field does not have a consensus on how much noise to deliver, which nerves to target, or how long the stimulation should be present.

\section{Future Work}
\par
Approaches to vibration therapy for neurorehabilitation share a common set of limitations that future studies need to address before the field can become a robust and reliable treatment method. Both whole-body vibration and local muscle vibration struggle with inconsistent stimulation parameters, unclear mechanisms, mixed results across different patient groups, and very limited long-term data. Study designs also vary widely, which makes it nearly impossible to compare results across trials or draw firm conclusions. Despite these variations, the available studies remain promising, showing enough short-term benefits to justify continued research.
\par 
Future work should focus on standardizing vibration parameters to better understand which specific changes in frequency, amplitude, or duration lead to meaningful improvements. More consistent goals across studies would also help clarify why these improvements occur, which is vital for building safe and effective clinical treatments. It will also be important for future studies to pay more attention to how factors like age, frailty, and neuromuscular status influence outcomes, since certain groups may naturally respond better than others. Ensuring more diverse and representative samples will make it easier to identify who benefits and why, so that we can understand which factors in health could be beneficial and which could pose a danger to high-risk patients. Larger study sizes, sham-controlled designs, and longer follow-up periods would help fix many of the current gaps. 
\par
Looking forward, wearable vibration devices offer one of the clearest paths for advancing the field. Wearables enable precise targeting \cite{seim2021wearable, Hung2015Wearable}, consistent dosing \cite{Wang2025MSVibration, Endo2024PosturalStability}, and home-based use \cite{Wang2022Wearable, Chandrashekhar2021DPN}, making them more practical for long-term rehabilitation than lab-based systems. Although they may not address all gaps, such as the need for more research into the impact of specific variables, they are essential because they are the most likely real-world applications of this technology. 
\par

\section{Conclusion}
\par
In this paper we discussed how vibration based interventions, ranging from whole body vibration platforms to resonant focused muscle vibration and stochastic wearable devices, modulate sensory and motor pathways to improve balance, gait, spasticity, and upper limb motor function in aging and neurologically impaired populations. Across these modalities, the review highlighted recurring challenges, including poorly standardized stimulation parameters, heterogeneity in protocols and outcome measures, small and clinically selective samples, and limited long term or mechanistic data, all of which complicate dose optimization, responder identification, and translation into routine care. At the same time, the surveyed studies demonstrate that, when paired with conventional therapy, vibration can yield clinically meaningful gains: whole body vibration supporting balance and mobility in selected groups, resonant focused muscle vibration reducing spasticity and enhancing motor coordination, and stochastic or wearable approaches improving dexterity and task specific upper limb use, while remaining relatively low burden and noninvasive. Future efforts should therefore prioritize rigorous, sham controlled trials with harmonized protocols, mechanistic endpoints, and more representative cohorts, while advancing wearable platforms that enable home based, adaptive, and user friendly delivery; together, these developments will be critical for turning vibration from a promising adjunct into a reliable, scalable component of comprehensive neurorehabilitation.

\bibliographystyle{ieeetr}
\bibliography{references}

\end{document}